\documentclass[conference]{IEEEtran}

\usepackage{cite}
\usepackage{amsmath,amssymb,amsfonts,amsthm}
\usepackage{algorithmic}
\usepackage{algorithm}
\usepackage{graphicx}
\usepackage{textcomp}
\usepackage{xcolor}
\usepackage{url}
\usepackage[inline]{enumitem}
\usepackage{verbatim}
\usepackage{listings}
\usepackage{multirow}
\usepackage{arydshln}
\usepackage{hyperref}
\newtheorem{definition}{Definition}

\begin{document}

\title{Fast Discovery of Inclusion Dependencies with Desbordante}

\date{}

\author{
\IEEEauthorblockN{Alexander Smirnov\textsuperscript{1}, Anton Chizhov\textsuperscript{1}, Ilya Shchuckin\textsuperscript{1}, Nikita Bobrov\textsuperscript{1}, George Chernishev\textsuperscript{1,2}}
\IEEEauthorblockA{\textsuperscript{1} Saint-Petersburg State University \\ 
\textsuperscript{2} Universe Data \\
Saint-Petersburg, Russia\\  \{alexander.a.smirnovv, anton.i.chizhov, shchuckinilya, nikita.v.bobrov, chernishev\}@gmail.com\\}}
\maketitle

\begin{abstract}

Inclusion dependency is a relation between attributes of tables that indicates possible Primary Key--Foreign Key references. Automatic discovery of inclusion dependencies is a relevant problem for both academic and industrial communities. The core concern for this problem is the efficiency of discovery process, since it is a computationally expensive task. However, existing studies only address the algorithmic side, while leaving out the implementation aspect. At the same time, engineering details are at least as important as the algorithmic ones for achieving good performance.

In this paper, we describe techniques for efficient implementation of two algorithms for discovery of inclusion dependencies~--- Spider and Faida. The first one is a classic algorithm whose ideas lie in the foundation of many other inclusion dependency discovery algorithms. We propose an efficient parallelization technique, which greatly speeds up the algorithm while simultaneously reducing its memory consumption. The second one is the state-of-the-art approximate algorithm, which we approach by applying four types of optimizations: data buffering, SIMD-enabled execution, careful hash-table selection and parallelization.

In order to experimentally evaluate our techniques, we have implemented these algorithms in Desbordante~--- an open-source science-intensive data profiler written in C++. For Spider, we have evaluated several different options, and in case of Faida we have demonstrated that all our optimization techniques yield results. We also compared our implementations with Metanome~--- a Java-based data profiler. Overall, we report up to 5x improvement in terms of run time reduction for Spider and up to 8x for Faida. 

\end{abstract}

\section{Introduction}\label{sec:introduction}

Inclusion dependency (henceforth referred to as IND) is one of the most well-studied database dependency concepts. First papers~\cite{CoddIND, SmithIND} on IND date back to the same epoch as studies of other classic dependencies like functional, multi-valued or join. Initially, IND was considered as a basis for defining a domain-key normal form, which guarantees that a relation has no insertion or deletion anomalies~\cite{FaginDKNF}. An IND complete axiomatization was soon proposed~\cite{FaginAxiomatization}, which defined the basis for all further work. 

In the past forty years IND has proved itself useful in data management tasks, such as query optimization~\cite{ModernQueryOptimizationFelix}, schema design~\cite{INDSchemaMatching}, and data integration~\cite{INDDataIntegration}. However, the IND is primarily used in foreign key discovery~\cite{INDForeignKeyDiscovery, INDForeignKeyDiscovery2}, which aims to suggest possible joins over tables.
In the first studies IND is also referenced as an \textit{interrelational constraint} since it is usually defined over a set of tables. It is worth mentioning that IND is not unique when it comes to multi-table setting. E.g., matching dependencies (MD)~\cite{MatchingDependencies} are also defined over a set of tables, although MD are primarily used for capturing common semantics over tables rather than discovering strong foreign key relationships.

Like any other database dependency type, we consider IND to be a knowledge that can be obtained in two different ways. First way is handled by an administrator or a user who possess a domain-specific knowledge. In that case, IND would be manually defined over a set of tables during the database design process. Such IND become a part of database integrity constraints that can not be violated, and for any table we can easily validate its constraints via database management system (DBMS) checks. However, it is very common for something as intrinsic as integrity constraints to be lost during data export, or to not have been defined in the first place. This is the use case for the second way of obtaining knowledge. We will refer to this way as ``automatic discovery of dependencies'', and the philosophy behind it is simple: ``if something was not defined for data, it can probably be mined from data''.
There is usually a high probability of such knowledge yielding useful insights.

Automatic discovery of dependencies is a well-studied problem from the algorithmic perspective, but not from the engineering one. In this paper, we develop a technical approach to the problem, which is based on efficient implementations of algorithms for IND discovery. Solving this problem can lead to more effective solutions of many data management tasks that employ IND. Therefore, the contribution of this paper is the following:
\begin{enumerate}
    \item A comprehensive study of Faida~\cite{KruseFastIND} and Spider~\cite{Spider}~--- two algorithms for automatic IND discovery, and designation of some of their components as candidates for optimization
    \item Implementation of several optimization techniques for Faida and Spider algorithms
    \item Experimental evaluation of the proposed optimizations
    \item Discussion of results and an outline of suggestions for effective implementation of two different types of algorithms for automatic IND discovery.
\end{enumerate}

In Sec.~\ref{sec:background} we provide the necessary context for our study. In Sec.~\ref{sec:relwork} we provide a short report for each IND discovery algorithm that can be found in modern papers on the subject. In Sec.~\ref{sec:algos} we scrutinize the two most important algorithms in this study, i.e. Faida and Spider. Next, in Sec.~\ref{sec:implementation} we discuss algorithm components that can be enhanced and propose appropriate optimization techniques. In Sec.~\ref{sec:experiments} we describe experimental setup, methodology, results, and takeouts. Threats to validity of this study are described in Sec.~\ref{sec:threats}. We conclude this paper with Sec.~\ref{sec:conclusion}.

\section{Background}\label{sec:background}

We provide IND definition according to the modern notation:

\begin{definition}
For two relations $R  = (R_1 \ldots R_k)$ and $S = (S_1 \ldots S_m)$, we define $\hat{R} = R_{i_1} \ldots R_{i_n}$ and $\hat{S} = S_{i_1} \ldots S_{i_n}$ as combinations of attributes of R and S respectively. We say that IND $\hat{R} \subseteq \hat{S}$ holds on relations R and S if for every tuple $t_r \in R$ there exists a tuple $t_s \in S$, such that $t_R[\hat{R}] = t_S[\hat{S}]$.
\end{definition}

One cay say that ``$\hat{R}$ is included in $\hat{S}$'', which explains the dependency type name. We call $\hat{R}$ the left hand side of a dependency, or \textit{dependent} columns, and $\hat{S}$ the right hand side of a dependency, or \textit{referenced} columns. If $\hat{R}$ and $\hat{S}$ are represented by \textit{a single attribute}, we call IND unary. If $\hat{R}$ and $\hat{S}$ are disjoint \textit{sets of attributes}, we call IND n-ary.
For example, we can say that relations presented in Table~\ref{tab:INDexample} comprise unary IND DLN $\subseteq$ DLID, since for any value of DLN attribute we can find the exact same value in DLID attribute.

\begin{table}[!h]
\centering
\vspace{-0.4cm}
\caption{IND: DLN $\subseteq$ DLID}
\label{tab:INDexample}
	\begin{minipage}{0.37\linewidth}
		\begin{tabular}{c|l|l|l|}
			\cline{2-4}
			$t_{id}$ & \textbf{UID}      &   \textbf{Name}    & \textbf{DLN}  \\ \cline{2-4}
            $t_{1}$ & 1        &   Sofia   & 21   \\ \cline{2-4}
            $t_{2}$ & 2        &   Leonard & 35   \\ \cline{2-4}
            $t_{3}$ & 3        &   Shavkat & 10   \\ \cline{2-4}
            $t_{4}$ & 4        &   Mary    & 65   \\ \cline{2-4}
            $t_{5}$ & 5        &   Andrew  & 10   \\ \cline{2-4}
		\end{tabular}
	\end{minipage}
	\hfill
    \hspace{0.01cm}
	\begin{minipage}{0.38\linewidth}
		\begin{tabular}{c|l|l|}
			\cline{2-3}
			$t_{id}$ & \textbf{DLID}      &   \textbf{Country} \\ \cline{2-3}
            $t_{1}$ & 21        &   Romania \\ \cline{2-3}
            $t_{2}$ &  35        &   Spain   \\ \cline{2-3}
            $t_{3}$ &  10        &   Germany \\ \cline{2-3}
            $t_{4}$ & 65        &   USA     \\ \cline{2-3}
		\end{tabular}
	\end{minipage}
\label{tabular:unaryIND}
\end{table}

Automatic discovery of IND represents the largest fraction of all contemporary studies on IND. We can categorize those studies by the type of IND mined:
\begin{itemize}
	\item unary IND~\cite{uIND, DeMarchi, Spider} or n-ary IND~\cite{Mind, ZigZag, Find2, Mind2}
	\item many tables setting~\cite{Many}
	\item approximate IND discovery~\cite{KruseFastIND, INDForeignKeyDiscovery2}, the approach which estimates the set of the actual INDs
	\item partial IND~\cite{LOPES20021, Mind}, which allows dependency violations by small fractions of typos or errors in data
	\item IND on non-relational data, e.g. on RDF~\cite{RDFind}
	\item distributed discovery~\cite{Sindy}
\end{itemize}

It can be seen that IND subject is researched in both deep and broad fashions, and all studies provide extensive evaluation of different aspects of automatic IND discovery.

Aforementioned studies keep pushing the envelope in terms of reducing run times, memory usage, and increasing volumes of data that can be processed by algorithms. Since algorithms are just high-level abstractions of ideas, any comparison study is free to choose the toolkit for algorithm implementation. 
Modern studies prefer to implement algorithms using tools and languages that prioritize ease of deployment and reproducibility. This approach leaves out potential improvements on the implementation side of things, leading to a lack of studies that are trying to solve engineering problems as opposed to algorithmic ones.

For example, the data profiling tool Metanome~\cite{Metanome} is a great playground that allows its users to get familiar with different database dependency concepts. It also provides an easy-to-use framework for developing dependency discovery algorithms. However, it is implemented in Java and thus can not guarantee the best run times or memory consumption rates. We believe that with the right implementation, the algorithms can be expressed in a more meaningful way, which would result in setting true industrial-level standards for automatic discovery tasks. That is the vision of our data profiling tool Desbordante~\cite{Desbordante}, which is responsible for the ``fast discovery'' part of this paper's title.

Desbordante is open-source, written entirely in C++, provides developers with core data structures necessary for dependency discovery tasks, and contains implementations of many well-known algorithms (\href{https://github.com/Mstrutov/Desbordante/}{github.com/Mstrutov/Desbordante}). It has both console and web versions, which can be tried here: \href{https://desbordante.unidata-platform.ru/}{desbordante.unidata-platform.ru}. It can be seen that in many ways Desbordante is inspired by Metanome~--- although the vision is quite different. Our goal lies in building not just a convenient tool to run research experiments, but an efficient, resilient to crashes, and scalable application for data mining. 

We want to emphasize, that in addition to most common database dependency concepts, Desbordante also provides its user with ways of obtaining or validating association rules, graph dependencies, algebraic constraints. Any type of knowledge that can be mined with Desbordante is called \textit{a primitive}~--- some hidden pattern in data. 

For this paper, we've implemented two algorithms in Desbordante that could mine the IND primitive: Faida~\cite{KruseFastIND} and Spider~\cite{Spider}. Initially, these algorithms were implemented within Metanome framework by authors of the experimental evaluation study~\cite{InclusionDependencyDiscovery}, and we used their Java code as the foundation for our work. We carefully converted the code to C++, preserving the initial logic of programs, and on top of C++ implementations we built the optimization techniques introduced in Sec.~\ref{sec:implementation}. The reasoning for choosing Faida and Spider as algorithms for our study is discussed in Sec.~\ref{sec:relwork-summary}.

\section{Related Work}~\label{sec:relwork}
In this section, we present the gist of each IND discovery algorithm developed over the past 40 years. We group them by the type of discovered IND as unary, n-ary, as well as both unary and n-ary. Due to the specifics of the n-ary IND discovery process, some of the algorithms require first $k$ levels of true INDs as an input for their mining process. We organize them into a stand-alone group in Sec.~\ref{sec:naryIND}. All the self-contained algorithms, which do not require an input and can mine IND starting with the very first level of search space, are listed in Sec.~\ref{sec:bothunIND}. In the last Sec.~\ref{sec:relwork-summary}, we analyze some significant characteristics of Faida and Spider algorithms.

\subsection{Unary IND algorithms}

\subsubsection{Bell and Brockhausen}
Bell and Brockhausen~\cite{BellAndBrockhausen} is a unary IND discovery algorithm that uses SQL-join statements for candidates validation. Authors use the transitivity of INDs to prune as many candidates as possible. However, this pruning strategy is not sufficient, since a large number of redundant candidates remain even after the filtering process.

Another issue with this algorithm is that it quickly becomes inefficient on large datasets due to the need to access the database for each candidate.

\subsubsection{DeMarchi}
The key idea of DeMarchi algorithm~\cite{DeMarchi} is to associate each value from a particular domain (the set of attribute's possible values) with every attribute containing this value. DeMarchi applies this data representation technique for candidate validation.
This method, while providing an efficient unary IND discovery, would have issues with most real-life applications due to large datasets, making such index too big to fit in the memory.

\subsubsection{Spider}
Spider~\cite{Spider} is a unary IND discovery algorithm. It solves the issue of inverted indexes becoming too large to fit in the memory by using disk space to store lists of values. It also achieved better efficiency than its predecessors by checking several candidates simultaneously and exploiting early termination for the candidate checks.

\subsubsection{S-indd}
S-indd~\cite{S-indd} is an extension of Spider that uses attribute clustering technique instead of an inverted index. This approach improves the scalability of the algorithm, making it independent of the number of attributes.

\subsubsection{Sindy}
Sindy\cite{Sindy} is a distributed unary IND discovery algorithm. This algorithm can be adapted to work with different map-reduce-style frameworks since its logic can be expressed in functional programming terms: map and reduce. By scaling the number of workers, Sindy can outperform most other methods. The distributed manner of working with datasets also eliminates the main memory constrain.

\subsubsection{Many}
Many algorithm~\cite{Many} is a unary inclusion dependency detection algorithm capable of working with large quantities of small tables. Web table corpus is a great example of such tables.

\subsubsection{S-indd++}
S-indd++~\cite{SinddPlusPlus} is a unary IND discovery algorithm that aims at eliminating drawbacks of S-indd and Binder. S-indd++ utilizes stepwise partitioning, which divides the dataset into buckets of different sizes. This method is capable of discarding a large number of attributes early on.

In contrast to Binder partitions are not required to fit into the main memory. This feature allows S-indd++ to avoid unnecessary operations of partition splitting which are computationally intensive.

\subsection{n-ary IND algorithms}\label{sec:naryIND}

\subsubsection{Mind}
Mind~\cite{Mind} is a n-ary IND discovery algorithm that uses Apriori-like candidate generation method. As input Mind requires precalculated uINDs. Authors use a separate algorithm for this purpose. Mind iteratively tests candidates of a certain size and generates new from the satisfied ones. Calculations stop when it cannot generate next candidates.

\subsubsection{ZigZag}
ZigZag~\cite{ZigZag} is an algorithm that uses a combination of pessimistic and optimistic approaches. Pessimistic approach uses adapted level-wise algorithm from Mind. Authors claim that level-wise approach is not efficient on large INDs. Instead, the algorithm constructs negative and optimistic positive borders and 'zigzags' between them. Negative and positive borders consist of all known smallest unsatisfied and largest satisfied INDs respectively. ZigZag uses distance estimation between positive border and satisfied INDs.

\subsubsection{Mind$_2$}
Mind$_2$~\cite{Mind2} is an n-ary IND discovery algorithm that uses the concept of unary IND coordinates. By using database queries, algorithm derives these coordinates from precalculated unary INDs. INDs can then be acquired by applying set operations on the coordinates.

\subsubsection{Find$_2$}
Find$_2$~\cite{Find2} is an algorithm that uses mapping between n-ary INDs discovery problem and 
Clique-Finding Problem in k-hypergraphs, which are graphs that have exactly k nodes connected by each edge.

The algorithm consists of two stages. At the first stage it discovers unary and binary INDs that will form the first hypergraph. Unary INDs form nodes of the graph, while k-ary INDs are represented by k-hyperedges.
At the second stage it calculates cliques using Hyperclique algorithm proposed by the authors of FIND$_2$. Additional validity checks of the discovered INDs are required since the clique property is necessary but not sufficient.

\subsection{Both unary and n-ary IND algorithms}\label{sec:bothunIND}

\subsubsection{Binder}
Binder~\cite{Binder} is an IND discovery algorithm that detects both unary and n-ary INDs. This algorithm was designed according to the Divide \& Conquer paradigm. Before processing, Binder splits datasets into smaller buckets and then checks these buckets for INDs. For validation, it uses two indexes: inverted and dense. A dense index reduces the number of candidate checks for a partition. Binder dynamically manages its memory. If buckets are too big to fit in memory, it splits them further. That greatly increases its scalability.

\subsubsection{Faida}
Faida~\cite{KruseFastIND} is an approximate IND discovery algorithm. 
Approximate algorithms are not guaranteed to produce correct or complete results. 
This assumption allows it to use more efficient discovery strategies thus significantly increasing performance. In particular, Faida provides complete yet incorrect results which can contain false positives. However, in~\cite{KruseFastIND} and in~\cite{InclusionDependencyDiscovery} was shown that Faida's false positive rate is quite small. In fact, it outperformed state of the art solutions of the time (such as Binder) without reporting any false positives.

\subsection{Summary}\label{sec:relwork-summary}
To conclude this section, we need to justify choosing Faida and Spider for our evaluation study.

Faida paper presents a one-of-a-kind approach, which addresses the problem of a trade-off between the correctness of discovered INDs and algorithm performance. Authors show that the proposed approach guarantees the output to contain the complete set (i.e. none are lost) of both unary and n-ary INDs, although it is prone to false positives. Faida is the state-of-the-art approximate IND discovery algorithm that demonstrates better performance on large datasets when compared to other IND discovery 
techniques~\cite{InclusionDependencyDiscovery} while maintaining an insignificant false positive rate. At the same time, the design of Faida guarantees better scalability~\cite{InclusionDependencyDiscovery} and does not require database connection for candidates validation subroutine as opposed to other n-ary IND discovery algorithms. Still, the architecture of Faida's solution contains many components that can be enhanced by using exclusively engineering solutions. All of that makes Faida a perfect candidate for our research, and it will be shown that Faida's components can be implemented in a more performant way.

Spider was the first unary IND discovery algorithm that utilized disk swap mechanism for solving the issue of data structures not fitting into RAM. This issue is the reason for prior algorithms falling short on datasets that produce large search space. We have selected Spider as the second algorithm for our research for multiple reasons. First, Spider is quite straightforward and can be easily implemented in C++. Second, unary INDs found by Spider are frequently used as an input for more advanced n-ary IND discovery algorithms~\cite{InclusionDependencyDiscovery}. Finally, Spider's core idea of backing up value files on disk is considered a solid foundation for many other algorithms~\cite{S-indd, Binder, Sindy, SinddPlusPlus}, so the proposed optimization techniques can also be applied to those algorithms.

\section{Algorithms}~\label{sec:algos}
In this section we will present a high-level overview of the algorithms considered in this paper.

\subsection{Faida}~\label{sec:faida}
The algorithm can be divided into three steps: preprocessing, candidate generation, and candidate validation.

\textbf{1. Preprocessing}. In this step the algorithm reads the input dataset in a per-table, line-by-line fashion. At this point the algorithm is solving two distinct problems~--- it converts the dataset into a representation better suited for further analysis and constructs a data sample for each table. The resulting entities will be used in further steps.

In order to construct the new representation, the algorithm splits each table into a number of columns, hashes each column's value and flushes results to disk. Each column is kept in a separate file (column file). In further steps the algorithm works exclusively with these hashed values, while the dataset itself is no longer in use.

We have to note that hashing, performed on this step, may result in collisions, which are one of the sources  of Faida's false positives.

In order to construct a data sample, the algorithm selects a subset of table rows that satisfies a specific condition. The sample must either contain all unique values for each column, or it must contain $n$ such values, in case a column contains more than $n$ of them. The value $n$ is the parameter of Faida which can be set up at launch time. This sample will be used for the construction of an inverted index, and therefore it is also hashed and stored on disk in a separate file.

\textbf{2. Candidate generation.} Faida starts by generating unary candidates, which constitute a set of all column pairs, excluding the ones where a single column appears twice. Then, candidate validation is run, and sets of n-ary candidates are formed. There is a partial order on this collection of candidates, so an algebraic lattice can be used to conveniently represent it.

Since the collection is rather large, the straightforward approach to generating candidates will require the algorithm to check too many candidates. To solve this issue, Faida uses Apriori-style generation, which is a popular approach used in several older algorithms. The candidate lattice is traversed in a level-by-level fashion, which allows the algorithm to eliminate many candidates that would not pass validation. Such traversal ensures that elimination happens without running a costly validation procedure.

Now, let us consider this part of the algorithm. The first level of lattice contains unary dependencies which were computed beforehand. This set is used to generate 2-ary candidates, which are then checked and the ones which pass validation form the resulting set of 2-ary valid INDs. Using this set, the 3rd level of lattice is computed and so on. Thus, the process is organized as follows: n-ary inclusion dependencies are used to generate (n+1)-ary candidates, which are subsequently validated to form a set of (n+1)-ary inclusion dependencies. It is repeated until the set of (n+1)-ary candidates is empty.

\textbf{3. Candidate validation}. For candidate validation Faida uses two data structures~--- an inverted index and the HyperLogLog. Let us consider them in detail.

The HyperLogLog (HLL)~\cite{hll} is a data structure that estimates the number of unique values in a set. Large datasets are handled well by this structure: it requires constant time and a constant amount of memory to compute the estimate. The original paper~\cite{KruseFastIND} states that if $X$ and $Y$ are columns or column combinations and $s(X)$, $s(Y)$ are sets of unique values in these columns, then $X \subseteq Y$ if and only if $|s(Y)| = |s(X) \cup s(Y)|$. HLL allows to estimate $|s(Y)|$ and $|s(X) \cup s(Y)|$, effectively allowing the validation of inclusion dependency $X \subseteq Y$. This data structure is parameterized by the accuracy of estimation. The higher the desired accuracy, the more memory is required.

However, the accuracy of the HLL suffers if the data contains a low number of unique items. For instance, this can occur in case of categorical columns. To handle this particular case, the authors opted to use inverted index. It is a data structure which maps a value into a set of columns containing that value. 

De Marchi et al.~\cite{DeMarchi} used an inverted index for the discovery of unary inclusion dependencies in the DeMarchi algorithm. The idea is to use the index to find all columns containing a specified one, and then intersect them. Applying this approach for each column makes it possible to discover all inclusion dependencies in the dataset.

However, it is too expensive for handling big datasets, since the inverted index becomes too large and does not fit into RAM. Therefore, Faida builds the inverted index from a table sample which was prepared on the preprocessing step. Note that since Faida hashes values, the inverted index uses hashes as keys as well.

The validation itself happens as follows. As input, this part accepts all candidates from the current lattice level. All data structures are initialized with hashed values that are read from the corresponding column files. Then, a candidate checking is performed. This process is based on the contents of the aforementioned structures. 

In the 2-ary, 3-ary, etc candidates, combinations of several attributes (e.g. $AB \subseteq CD$) start to appear in left and right parts of INDs. Because of that, Faida has to check set containment for tuples instead of single values. In the straightforward approach, it would have to fill data structures with tuple data. However, as Faida works with hashed data, a different approach is taken: each sequence of hashes (corresponding to a tuple) is combined into a single hash using XOR and circular shift operators. Thus, even for n-ary candidates the validation process is in no way different from the unary case.

\subsection{Spider}

Similarly to Faida, Spider consists of the same three steps: preprocessing, candidate generation, and candidate validation.

\textbf{1. Preprocessing.} At this step Spider iterates over input tables. Each column is sorted, deduplicated, and then flushed to disk into a separate file (column file). During this step the algorithm can write to disk if the program approaches memory limit. In that case, the algorithm processes data in parts, similarly to external sort in DBMSes.

\textbf{2. Candidate generation.} An attribute object is an object created for each attribute. It contains:
\begin{itemize}
    \item a forward iterator for enumerating column data for the corresponding attribute;
    \item two lists of attributes~--- a list of referenced and a list of dependent attributes.
\end{itemize}

Thus, each attribute object stores information on which attributes may be present in left and right parts of the prospective dependency. In the former case object's attribute is the referenced one, and in the latter case it is the dependent one.

At this step, Spider initializes all attribute objects with iterators pointing to the first value of the corresponding column file. Both lists are populated with all attributes except for the current one. An advanced version of the algorithm may filter these lists using column data type (e.g. no point in checking dependency between float and integer strings). The basic version (which is considered in this paper) treats all columns as the string ones.

Attribute object is considered \underline{processed} if its column has been fully read or both lists are empty.

\textbf{3. Candidate validation.}

Attribute objects are stored in the min-heap, sorted according to the values pointed to by their iterator. Spider repeats the following process until min-heap is not empty:
\begin{enumerate}
    \item A set of attribute objects sharing the same value is extracted from the min-heap. We denote it as the \underline{set}.
    \item For each attribute object belonging to the \underline{set}, its referenced attributes are intersected with the \underline{set}. In other words, the algorithm removes attributes that are not included in the \underline{set} from the referenced list. Dependent attributes are removed in a similar manner.
    \item For each \underline{unprocessed} attribute object from the \underline{set}, the algorithm increments iterator and inserts the attribute object into min-heap.
\end{enumerate}

\section{Implementation}\label{sec:implementation}

\subsection{Faida}\label{sec:implementation-faida}

Having studied the performance of the algorithm, we have discovered that the majority of time spent was on filling in data structures. This process is a part of the candidate validation step (recall Sec.~\ref{sec:faida}). At the same time, candidate checking itself, which is also a part of the validation step, had almost no contribution to the total run time.

Let us consider the algorithm part that is responsible for filling in data structures for each candidate level.
In the outer loop the algorithm iterates over table rows, and for each row it reads values from files containing hashed columns. This way, it constructs one dimensional array containing hashed representation of the original table row. In every iteration, the algorithm also iterates over column combinations on the current level, and for each one of them it computes a combined hash. That is, it computes a hash for a row, projected (in relational database terms) on the column combination. This hash is then added into the inverted index or the HLL. If this hash is already among keys of the inverted index, then we add the current column combination to the list of column combinations that corresponds with this key. Otherwise, we add this hash into the HLL, since the current column combination is not covered by the inverted index.

\textbf{Buffering.} The main drawback of this approach is that we have to process each table row individually, which is an inefficient way of using CPU. In order to address this issue, we have proposed the use of data buffering. The simplified version of the modified algorithm is shown in Listing~\ref{listing:faidaBuffering}. Its inputs are: \textit{table}, \textit{inverted index}, and \textit{levelColumnCombs} structure. The latter stores all column combinations belonging to the current candidate lattice level and, apart from that, maps these combinations into the corresponding HLL data structures.

Now, the outer loop iterates not over individual table rows, but over chunks~--- a series of consecutive rows. Therefore, we have to read not a single value from each involved column file, but a block of values (lines 3-4) in order to obtain data for several rows. 

The data is kept in a \textit{hashedChunk}, which is a two-dimensional array with the first index denoting the column number, and the second one representing the row number. This allows us to process whole chunks instead of individual rows. First, we calculate combined hashes for rows projected on the current column combination (lines 8-10) and then insert these values into the inverted index and the HLL (lines 11-17). This approach should reduce the number of cache misses since 1) data is densely packed into an array, and 2) method calls exhibit better temporal locality.

\begin{algorithm}
  \small
  \caption{Faida~--- filling data structures, buffered}\label{listing:faidaBuffering}
  \begin{algorithmic}[1]
    \REQUIRE{$table, \mathit{invertedIndex}, \mathit{levelColumnCombs}$}
    \FOR{$chunk$ in $table$}
        \STATE $\mathit{hashedChunk}\gets \mathit{ArrayOfIntegers2D}(\newline \hspace*{2em} table.\mathit{numColumns}, 
     \mathit{chunkSize}\newline \hspace*{1em})$

        \FOR{$\mathit{colFile}$ in $table.\mathit{colFiles}$}
            \STATE $\mathit{hashedChunk}[i]\gets \newline \hspace*{1em}\mathit{ReadNextBlock}(\mathit{colFile}, \mathit{chunkSize})$
        \ENDFOR

        \FOR{$(\mathit{columnCombination}, hll)$ in $\mathit{levelColumnCombs}$}
        \STATE $\mathit{combinedHashes}\gets \mathit{ArrayOfInts}(\mathit{chunkSize})$

        \FOR{$\mathit{columnIdx}$ in $\mathit{columnCombination}$}
            \STATE $\mathit{combinedHashes}\gets \mathit{RotlXor} (\newline
            \hspace*{2em} \mathit{combinedHashes},  \mathit{hashedChunk}[\mathit{columnIdx}]\newline
            \hspace*{1em})$
        \ENDFOR
        \FOR{$hash$ in $\mathit{combinedHashes}$}
            \IF {$\mathit{invertedIndex}.\mathit{HasKey}(hash)$}
                \STATE $\mathit{invertedIndex}.Insert(\newline 
                \hspace*{2em} hash, \mathit{columnCombination}\newline
                \hspace*{1em})$
            \ELSE
                \STATE $hll.Insert(hash)$
            \ENDIF
        \ENDFOR
        
        \ENDFOR
    \ENDFOR
  \end{algorithmic}
\end{algorithm}

\textbf{Hash table.} Note that there are two operations involving the inverted index (lines 12-13): a lookup of hashed value and an addition of new columns that contain this hash. In Metanome, the inverted index was implemented as a hash table with integer keys. It maps hashed values onto sets of column combinations that contain these values. Individual column combinations are encoded as integer values, and therefore are represented by sets of integers and implemented as hash tables. That is why Metanome doesn't use default Java hash tables, instead opting for implementations from the \verb+fastutil+ library, which are optimized for integer keys.

Our initial experiments have demonstrated that hash table from the standard C++ library failed to achieve the desired level of performance. For this reason we have also decided to use third-party implementation. There are lots of various hash table implementations for C++ with an extensive comparison of available options (\href{https://martin.ankerl.com/2022/08/27/hashmap-bench-01/}{martin.ankerl.com/2022/08/27/hashmap-bench-01}).

In the considered code the lookup speed is a priority: a dataset having $N$ rows and $M$ column combinations will issue $N*M$ inverted index lookups. Therefore, we have selected the \verb+emhash7+ hash table, a member of the \verb+emhash+ family (\href{https://github.com/ktprime/emhash}{github.com/ktprime/emhash}). It is optimized for storing integers, and according to the benchmarks offers fast lookup speed. We have also selected the \verb+emhash+ family hash table (\verb+emhash2+) with a high performance for inserts to represent sets of column combinations.

\textbf{Vectorization.} Now, consider Listing~\ref{listing:faidaBuffering}, line 9. Note that the process of computing combined hashes for projected rows is presented in the vectorized form: ROTL (left circular shift) and XOR are performed not for individual hashes, but for the whole chunk. Note that the circular shift is performed by one bit to the left, as in the original implementation.

A straightforward implementation of \verb+RotlXor+ function will iterate over an array of hashes and call ROTL and XOR instructions for each element. Therefore, we have decided to vectorize these computations using SIMD instructions.

\begin{algorithm}
  \small
  \caption{RotlXor Vectorized}\label{listing:RotXorImproved}
  \begin{algorithmic}[1]
  \REQUIRE{$\mathit{combinedHashes}, \mathit{hashedChunk}[\mathit{colIdx}]$}
  \STATE $\mathit{rowIdx} \gets 0$
  \FOR{$\mathit{hashVector}$ in $\mathit{combinedHashes}$}
    \STATE $\mathit{vectReg} \gets \mathit{LoadVect}(\mathit{hashVector})$
    \STATE $\mathit{vectReg} \gets \mathit{RotlVect}(\mathit{vectReg})$
    \STATE $\mathit{vectReg} \gets \mathit{XorVect}(\newline \hspace*{2em} \mathit{vectReg}, \mathit{hashedChunk}[\mathit{colIdx}][\mathit{rowIdx}]\newline \hspace*{1em})$
    \STATE $\mathit{hashVector} \gets \mathit{StoreVect}(\mathit{vectReg})$
    \STATE $\mathit{rowIdx} \gets \mathit{rowIdx} + \mathit{vectorSize}$
  \ENDFOR
  \end{algorithmic}
\end{algorithm}

Now, several items of \textit{combinedHashes} are loaded into vector register (lines 2-3 of Listing~\ref{listing:RotXorImproved}) and vectorized versions of ROTL and XOR operations are executed. Thus, the technique uses the same number of instructions as the straightforward approach to compute combined hash for several rows. The obtained speedup depends on the number of rows that can fit into the register. Our implementation is tailored to 64-bit integers and the AVX2 instruction set, which offers 256-bit registers. Thus, our technique allows the computation of four hashes with a single instruction.

Unfortunately, vector extensions up to AVX2 (including AVX2 itself) do not support vectorized ROTL. Therefore, we had to emulate it using three other vectorized instructions: two bitwise shifts and a bitwise or. The idea is as follows: the original number is 1) shifted $1$ bit left, 2) shifted $n-1$ bits right, where $n$ is hash size in bytes, 3) results of the first and second step are ``or''ed. 
Thus, in our implementation two regular instructions will be substituted with four vectorized ones: three for ROTL and one for XOR. Finally, it should be noted that the next version of vectorized instruction set (AVX-512) offers a single vectorized instruction ROTL. This will further improve the performance of this part of the algorithm.

\textbf{Parallelization.} Despite various optimizations, the process of filling in data structures for big datasets remains a bottleneck.
Let's consider a for loop at lines 6--18 in Listing~\ref{listing:faidaBuffering}.
For each chunk, this loop iterates over each pair of column combinations and the corresponding HLL. 
In each iteration, it performs combined hashes calculation (lines 8--10) and a data structure insertion (lines 11--16).
According to the profiling results, this loop significantly affects performance, which is why we decided to parallelize it.

Since every column combination can be encoded with an integer, the inverted index is represented with \verb+std::map<int, std::set<int>>+, which maps a hash to the set of columns containing that hash.
Due to this design, a race condition can arise if two threads were to acquire equal hashes and attempt to perform the insertion operation (\verb+invertedIndex.Insert()+) on a corresponding set.
There are two possible ways to tackle this problem.
The first one is to create a mutex (or a spinlock) for each \verb+std::set<int>+, thus preventing simultaneous column combinations insertions.
The second one is to replace \verb+std::set<int>+ with a lock-free bit vector.
In this case, setting the i-th bit is equivalent to inserting the i-th combination in a set.
This operation is atomic, preventing the aforementioned race conditions from occurring.
The advantage of this approach is the absence of system calls that appear when working with mutexes. 
On the other hand, it increases memory consumption since every bit vector must be allocated beforehand to reserve space for every combination, while the first method initially has empty int sets.

\subsection{Spider}\label{sec:implementation-spider}Early experiments with Spider have demonstrated that the most time- and memory-consuming step was the preprocessing. At the same time, candidate generation and validation steps take an order of magnitude less time. Overall, their contribution to the total run time is negligible and therefore will not be considered in our study. Instead, our enhancements concern only the first step of the algorithm, for which we describe several low-level techniques that present a trade-off between run time and memory consumption.

In Metanome, the preprocessing step is performed on a column-by-column basis, i.e. for each column, the algorithm performs a full table scan, and writes sorted and deduplicated values into a separate file. To store deduplicated values, Metanome uses a sorted set of unique elements represented by a \verb+java.util.TreeSet+ data structure. While processing large datasets, the Spider may flush intermediate results to disk when hitting the memory limit. In such cases, it has to merge these intermediates later. On the one hand, Metanome's approach is quite simple implementation-wise and is memory-efficient. But on the other hand, reading the source table multiple times may lead to subpar performance. Now, let us turn to our implementations.

\textbf{Set-based implementation.} Our first implementation uses the \verb+std::set+ parameterized with \verb+std::string+ to store intermediates. It is a single-pass implementation, which flushes intermediates for all attributes to disk when the memory limit has been reached. The set-based approach eliminates duplicates for free, but creates a potential bottleneck:  set insertion will take a lot of time. Moreover, an efficient multi-threaded execution will not be possible. As the result, both Metanome's Spider and our set-based implementation are single-threaded applications.

\textbf{Vector-based implementation.} In order to ensure the parallelizability of the algorithm, we have devised a second implementation, which uses \verb+std::vector+ to store results. In this approach, vectors are initially populated with values from the corresponding columns. Then, vector sorting and duplicate elimination are run in parallel, with each vector being processed in a separate thread. This way a significant speedup can be achieved.

The main drawback of this approach is the amount of used memory since vector stores all duplicate values. Therefore, it is essential to reduce the amount of memory needed for storing results. A straightforward approach of using \verb+std::string+ as a vector key is a very inefficient option, since each item takes at least 32 bytes. As the result, the original table will grow several times.

To make the implementation more efficient memory-wise, we have implemented a chunk-based reading of the source files. Each file is split into equal-sized parts, such that each part can not occupy more than half of available memory. The remaining memory is used to store column intermediates. Thus, loading and processing tables in parts resemble an external sort operation in DBMSes. Consequently, at the end of this processing, the algorithm will have to merge resulting intermediates and build a single output file.

The chunk-based processing leads to vector storing references instead of actual values, with those references being tied to a position relative to the current part. There are two possible implementations which use different keys:
\begin{enumerate}
    \item \verb+std::string_view+~--- values of this type reference a contiguous region of memory.
    \item \verb+std::pair<uint, uint>+~--- the first element stores an offset from the start the part, and the second stores its length. In this case the maximum part size is capped by maximum integer value.
\end{enumerate}

As we noted earlier, the code of candidate generation and validation is similar to the one used in Metanome. However, we would like to mention a small optimization that is possible in our approach and that may further increase the performance. It is possible to add checks for maximum values that will run during the attribute object initialization. Observe the following simple heuristic: inclusion dependency $A \subseteq B$ will not hold if the maximum value in A is less than the maximum  value in B. Checking this fact while populating both lists will allow the algorithm to reduce the number of iterations during the candidate validation step.

\section{Experiments and Discussion}\label{sec:experiments}
In this section we present an experimental comparison between our implementations of the aforementioned algorithms and their Java counterparts from Metanome profiler.

\subsection{Experimental Setup}

\textbf{Methodology.}

In Sec.~\ref{sec:algos} it was shown that Faida and Spider workflows share the same three stages, but they differ in the data structures, as well as subroutines and discovery methods. Due to the specifics of algorithms, we address different optimization techniques, which is why we provide detailed descriptions of experiments in the corresponding sections: Sec.~\ref{sec:experiments-faida} for Faida and Sec.~\ref{sec:experiments-spider} for Spider. In this section we focus on the high-level methodology related to both algorithms. 

The methodology of our evaluation study is defined by two research questions:

\begin{enumerate}
	\item \textit{RQ1: which families of optimization techniques can be considered as the more efficient for implementing automatic IND discovery algorithms?}
	\item \textit{RQ2: how significantly can an engineering approach to the problem of automatic IND discovery improve state-of-the-art solutions?}
\end{enumerate}

To answer those research questions, we implemented Faida and Spider for Desbordante in C++. Both implementations are based on the Metanome versions of algorithms written in Java.
For both algorithms \textit{run time} is used as the main metric for performance assessment. We consider average execution time metric, which is acquired by averaging measurements of 5 runs. The disk cache is cleared on each run to avoid skewing the results. Lastly, we handle \verb+NULL+ values according to the following well-known policy: \verb+NULL+ values are equal to each other, but can not be treated as ``synonyms'' to any other value.

For Spider we also measured \textit{memory consumption} since it is the other most prominent aspect of applied optimizations for this algorithm.

For each experiment, we provide a review of obtained results as well as the impact on performance as the optimization techniques are applied. Discussion of these results forms the answer to \textit{RQ1}. To answer \textit{RQ2}, we compare our best implementations of Faida and Spider with the Metanome ones.

\textbf{Datasets.} 
For evaluation, we use 3 real datasets from different domains and 3 synthetic ones. The datasets with their characteristics are listed in Table~\ref{tab:datasets}. We aimed to construct a collection of datasets such that each of them would possess some unique trait. \textit{Brazilian E-commerce} (\textit{ECOMM} for short) is a real sales dataset which contains rather long strings with user feedback. \textit{FitBit Fitness Tracker Data} (\textit{FITBIT}) is a wide dataset which holds a great number of INDs. \textit{CIPublicHighway (CI\_PH)} is a  sparse dataset with big percentage of \verb+NULL+s. \textit{TPC-H} is a widely used synthetic dataset for which we consider two scale factors: 1 and 10. The last dataset \textit{haINDgen} is generated by an utility which can produce data with n-ary INDs, where n is large.

\begin{table*}[!h]
    \centering
    \vspace{-0.4cm}
    \caption{Datasets characteristics}
    \label{tab:datasets}

    \begin{tabular}{l c c c c c c c c c}
         \hline
         \textbf{Dataset} & \textbf{Size} & \textbf{Type} & \textbf{Tables} & \textbf{Attributes} & \textbf{Rows}  & \textbf{Unary INDs} & \textbf{n-ary INDs} & \textbf{Max nIND arity} \\ \hline \hline
         Brazilian E-Commerce~\cite{olist}    & 126.3 MB & Real-world & 9 & 52 & 1.6 M  & 23 & 23 & 1\\ 
         FitBit Fitness Tracker Data & 330 MB & Real-world  & 18 & 259 & 8.1 M  & 8798 & ? (ML) & ? (ML)  \\ 
         TPC-H (SF=1)                & 1.1 GB & Synthetic  & 8 & 61 & 8.7 M  & 96 & 99 & 2\\ 
         TPC-H (SF=10)               & 11.1 GB & Synthetic  & 8 & 61 & 86.6 M  & 97 & 103 & 3\\ 
         haINDgen (generator)        & 862 MB & Synthetic  & 2 & 36 & 6.1 M  & 18 & 221 & 7\\ 
         CIPublicHighway                    & 28.3 MB & Real-world & 1 & 18 & 0.4 M  & 65 & 440 & 4\\
         \hline
    \end{tabular} 

\end{table*}

\textbf{Hardware and software.}
For experiments we used a PC with the following specs: AMD Ryzen 7 4800H CPU @ 2.90GHz x 8 cores (16 threads), supports AVX2 vector extension, 32 GiB RAM, SSD A-Data S11 Pro AGAMMIXS11P-512GT-C 512GB, running Ubuntu 22.04.

Java configuration used by Metanome:
openjdk 19.0.1 2022-10-18
OpenJDK Runtime Environment (build 19.0.1+10-Ubuntu-1ubuntu122.04)
OpenJDK 64-Bit Server VM (build 19.0.1+10-Ubuntu-1ubuntu122.04, mixed mode, sharing).

For Desbordante, the following software was used: gcc (Ubuntu 11.3.0-1ubuntu1~22.04) 11.3.0, ldd (Ubuntu GLIBC 2.35-0ubuntu3.1) 2.35. Our implementations were compiled with -O3 option.

\subsection{Experiment 1: Faida}\label{sec:experiments-faida}

\begin{table}[!h]
    \vspace{-0.4cm}
    \caption{Faida~--- Overall speedup}
    \label{tab:faida-all-speedup}
    \setlength{\tabcolsep}{3pt}

    \resizebox{\columnwidth}{!}{
    \begin{tabular}{rcccccc}
    \hline
    & CIPH & FIT(U) & HAIND & ECOMM & TPCH-1 & TPCH-10 \\
    \hline \hline
    Sequential(5) & 2.60 & 1.65 & 3.14 & 2.21 & 2.52 & 2.52 \\
    Parallel(6) & 3.68 & 1.61 & 8.20 & 2.27 & 3.40 & 3.35 \\
    \hline
    \end{tabular}
    }
\end{table}

\begin{table}[!h]
    \vspace{-0.4cm}
    \caption{Faida~--- Filling data structures speedup}
    \label{tab:faida-fill-speedup}
    \setlength{\tabcolsep}{3pt}

    \resizebox{\columnwidth}{!}{
    \begin{tabular}{rcccccc}
    \hline
    & CIPH & FIT(U) & HAIND & ECOMM & TPCH-1 & TPCH-10 \\
    \hline \hline
    Sequential(5) & 2.64 & 2.66 & 3.22 & 2.48 & 3.11 & 3.27 \\
    Parallel(6) & 5.47 & 1.92 & 10.20 & 3.64 & 8.14 & 9.11 \\
    \hline
    \end{tabular}
    }
\end{table}

\textbf{Experiment.}
In order to demonstrate that each of the proposed techniques increases the resulting performance, we have evaluated five implementations that incrementally introduce discussed optimizations. The first one, \underline{Default(2)} is the C++ equivalent of the \underline{Metanome(1)}. It is a straightforward rewrite that does not contain any optimizations. The first non-trivial version \underline{HTab(3)} features a specialized hash table. Next, \underline{HTab+Buff(4)} and \underline{HTab+Buff+SIMD(5)} versions contain additional buffering and buffering with SIMD-enabled computations, respectively.

Finally, \underline{HTab+Buff+SIMD+Par(6)} is a parallelized version of the \underline{HTab+Buff+SIMD(5)} implementation. The parallel version was run using 16 threads. Recall that earlier (see Sec.~\ref{sec:implementation-faida}) we have discussed two parallel implementations. Here, we benchmark only the mutex-based one. This was done to provide a more fair comparison since, unlike the lock-free version, the mutex-based version requires the same amount of memory. 

In Sec.~\ref{sec:faida} we noted that Faida has two parameters. The first one defines a sample size, and the second is used to determine the accuracy of HyperLogLog. These parameters impact not only the accuracy of the algorithm but also its run times. In our experiments, we have decided to use the same parameter values as in experiments in the original paper~--- sample size was set to 500 and HLL accuracy to 0.1\%. Authors of the original paper~\cite{KruseFastIND} experimentally demonstrated that these values ensure a sufficient quality of results without causing a significant slowdown.

\begin{figure*}[h]
	\centering
    \includegraphics[width=0.8\textwidth]{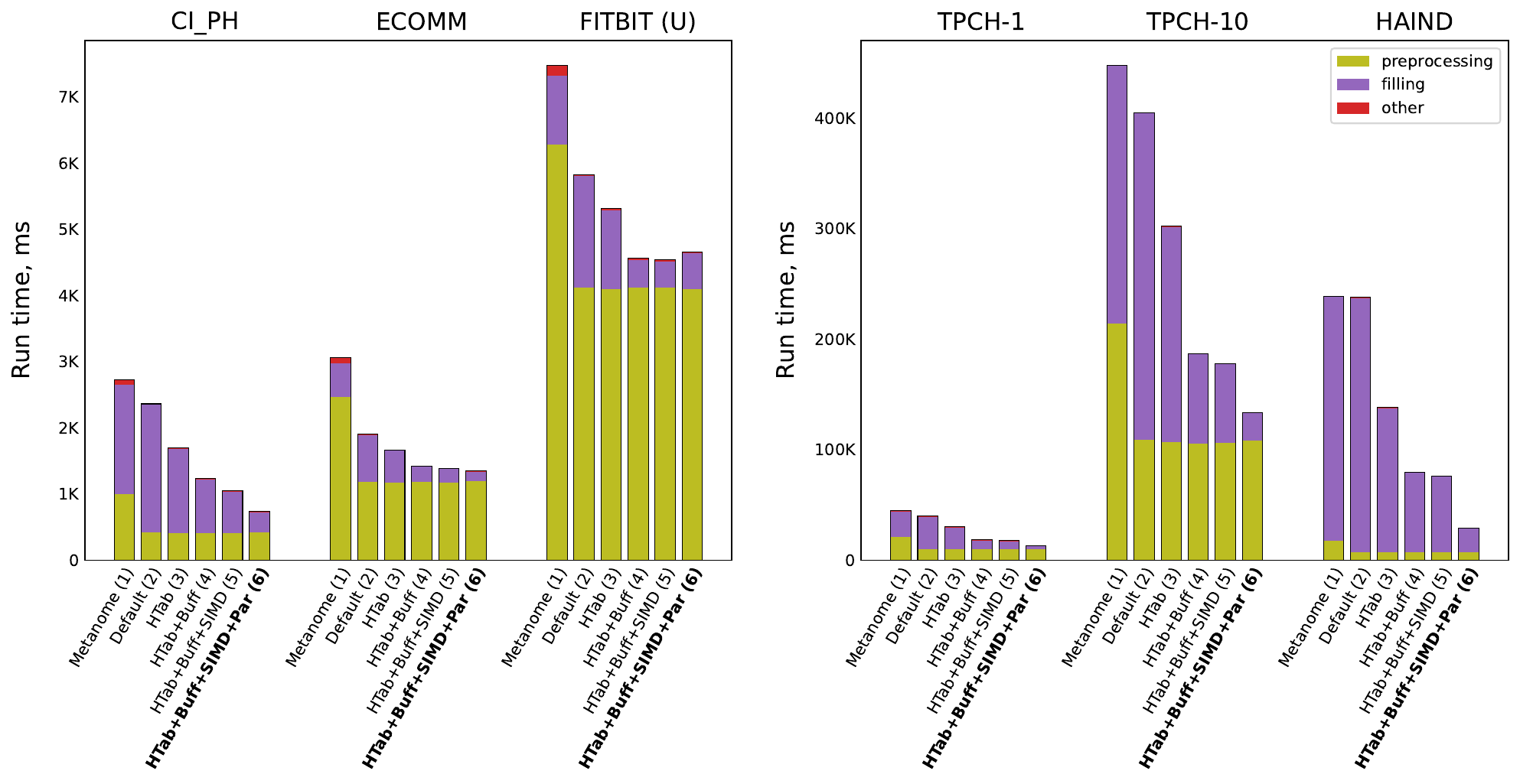}
	\caption{Run time of implementations of Faida on different datasets.}
	\label{fig:faida}
\end{figure*}

For each of the implementations, we have measured average run times while separately recording the preprocessing and population phase times. Results are presented in Fig.~\ref{fig:faida}. Each figure is as follows: the datasets are listed above, and for each of them the implementations are presented below. Note that we have to divide our collection of datasets into two figures for better presentation as their run times vary significantly.

Another point of concern is the accuracy of Faida, since it is an approximate algorithm. In our experiments, we do not evaluate it because it was already studied in the earlier works~\cite{KruseFastIND, InclusionDependencyDiscovery} and none of our techniques should impact it.

\textbf{Results \& Discussion.} Experiments demonstrated that each of the proposed optimizations gave a positive result, speeding up the algorithm. We can note that using a special hash table (\underline{HTab(3)}) and buffering (\underline{HTab+Buff(4)}) provided relatively better speedup than SIMD computations (\underline{HTab+Buff+SIMD(5)}). However, all of these optimizations yield improvement, therefore the \underline{HTab+Buff+SIMD(5)} version can be recommended in a single-threaded environment. The overall best performance was demonstrated by the \underline{HTab+Buff+SIMD+Par(6)} version, which we recommend for use in a multi-core environment. The obtained speedups (compared to Metanome) are provided in Table~\ref{tab:faida-all-speedup}.

Interestingly, the FITBIT dataset cannot be processed with Desbordante and Metanome implementations. There are more than 5 million of 2-ary dependencies alone, so during candidate generation for the third level both implementations run out of memory. Therefore, we provide the numbers only for unary dependencies in this case (hence, the U mark in the figure).

Next, consider algorithm run times during the preprocessing and filling phases. Our optimizations involved only the latter and significantly speeded it up for all datasets. The obtained speedups (compared to Metanome) for both best sequential and parallel implementations are presented in Tables~\ref{tab:faida-fill-speedup}. The improvement depends on various dataset properties, such as the number of dependencies, the number of unique values, and others. It should be also mentioned that the performance of this phase can be affected by disk speed as Faida reads column files.

Another artifact that we need to discuss is the performance of parallel and sequential implementations on the FITBIT dataset. There, the parallel implementation demonstrated worse results than the sequential version. We believe that the reason for this lies in the specifics of the dataset: FITBIT contains many identical values which leads to many mutex lock/unlock events during the inverted index filling. We have also evaluated the lock-free implementation which we described above, and in that case, no performance degradation occurred.

Now let's talk about the preprocessing phase (see Sec~\ref{sec:faida}). In our experiments, its run time is constant for a fixed dataset regardless of the implementation since we have not optimized it. This phase is rather straightforward and depends on three aspects: data parser, disk read/write speeds, and a hash function. Our implementation uses the internal Desbordante parser, which is why we have not tried to speed it up or parallelize it. However, we have experimented with an efficient third-party parser and found out that it was possible to obtain a significant speedup. Disk speed also heavily impacts the performance of this phase: the algorithm not only reads an input file, but also writes column files at the same time. The hash function is the last aspect that has to be carefully taken into account. A good hash function can be slow, but its quality determines the collision frequency, which impacts the accuracy of the overall algorithm.

Interestingly, candidate generation and checking phases provide a negligible contribution to the overall run time and cannot be even seen in the resulting figures, except for the FITBIT dataset. We have studied this issue, and we believe that generation and validation will be noticeable if there are a lot of candidates and dependencies, like in the FITBIT case with n-ary INDs.

\begin{figure*}[h]
	\centering
    \includegraphics[width=0.8\textwidth]{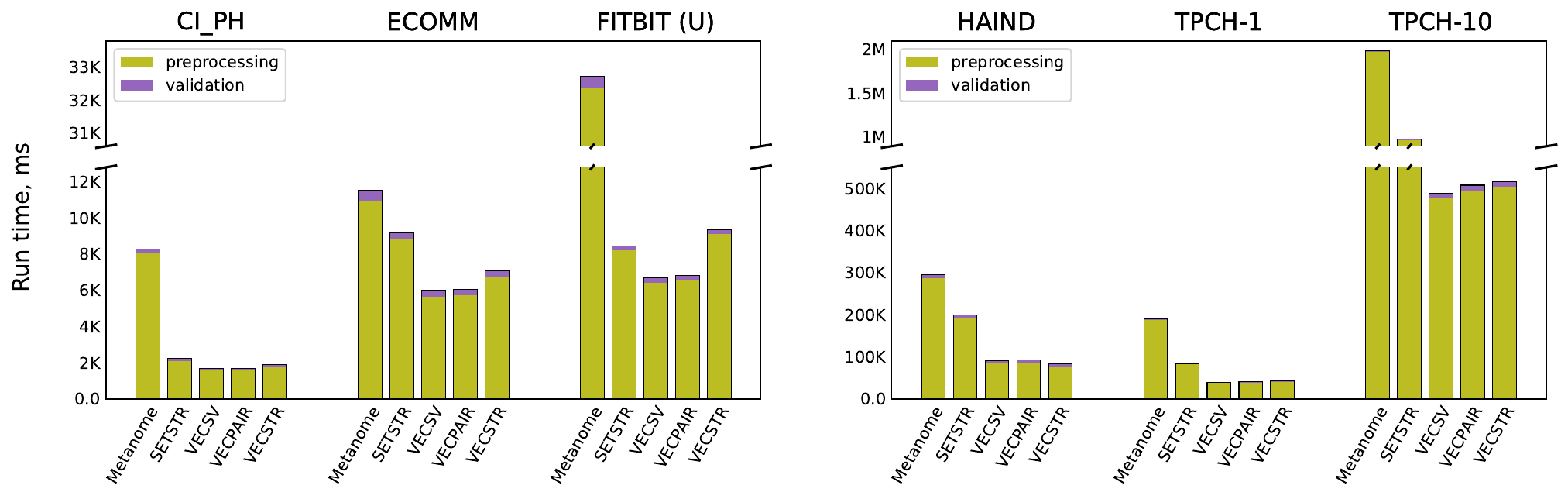}
	\caption{Run time of implementations of Spider on different datasets.
 }
	\label{fig:spider-runtimes}
\end{figure*}

\begin{figure*}[h]
	\centering
    \includegraphics[width=0.8\textwidth]{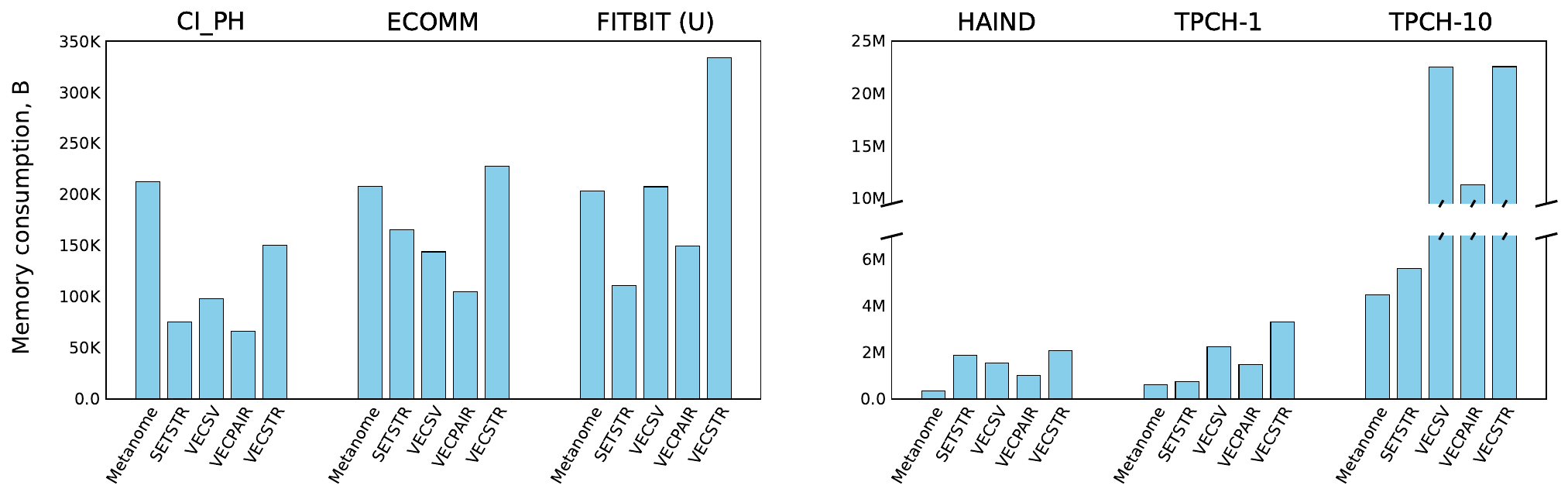}
	\caption{Memory consumption of implementations of Spider on different datasets.
 }
	\label{fig:spider-memory}
\end{figure*}

\begin{figure*}[h]
	\centering
    \includegraphics[width=0.8\textwidth]{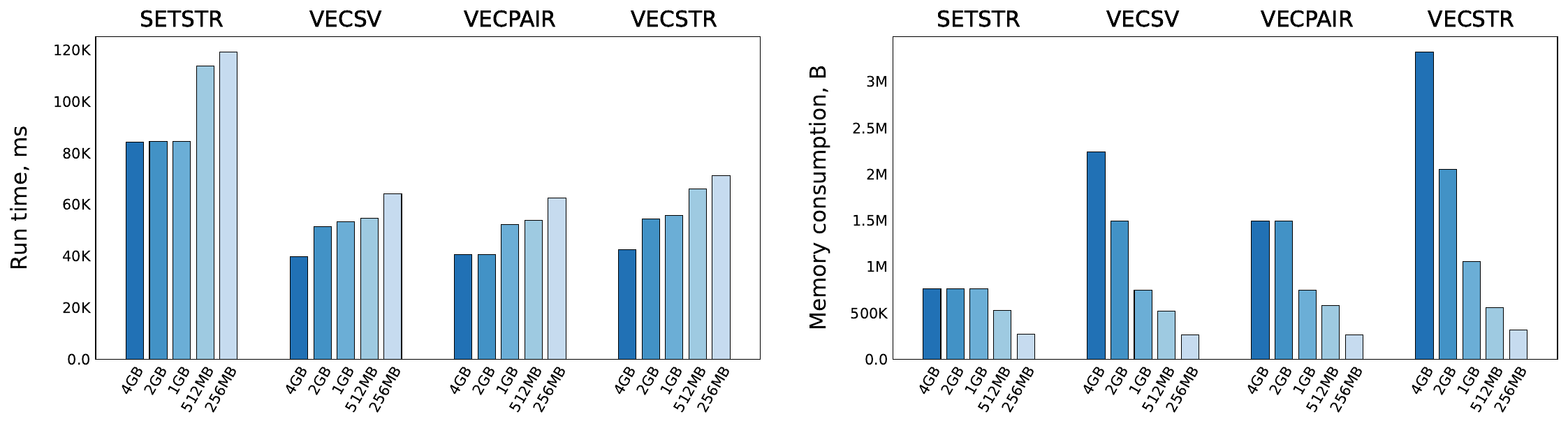}
	\caption{Time and memory consumption of Spider on Varying Memory Limits.
 }
	\label{fig:spider-experiment}
\end{figure*}

\subsection{Experiment 2: Spider}\label{sec:experiments-spider}
\textbf{Experiments.} In this section, we will use the following names to denote different implementations:
\begin{enumerate}
    \item SETSTR and VECSTR denote implementations that employ \verb+std::string+
    \item VECSV and VECPAIR denote implementations that rely on \verb+std::string_view+ and \verb+std::pair<uint, uint>+ respectively
\end{enumerate}

We have created all four implementations that we discussed in Sec~\ref{sec:implementation-spider}. The VECSTR was included into the evaluation pool in order to demonstrate the excessive memory footprint of this approach.

The run time and memory consumption are the primary metrics for the Spider algorithm and are assessed in our first experiment. The results are presented in Fig.~\ref{fig:spider-runtimes} and Fig.~\ref{fig:spider-memory}. The \verb+std::vector+-based implementations were evaluated with the maximum number of threads (16) with the memory limit set to 22GB.

The Metanome evaluation included the selection of the optimal values for the JVM -Xmx parameter. It is used to define the maximum amount of memory heap for JVM. This way, we ensure that the algorithm never flushes intermediate results to a disk. If this parameter is not used, then the Spider consumes much more memory, since Garbage Collector almost never frees heap until the end of processing. Finally, we tried to ensure that run times of Spider were affected only slightly or not at all when we were selecting this value.

In the Metanome implementation, candidate generation step is inseparable from the preprocessing. Therefore, in the Fig.~\ref{fig:spider-runtimes} preprocessing includes both of these steps. These steps can be merged into one, since, as we noted in Sec.~\ref{sec:implementation-spider}, the candidate generation contribution to the total time is negligible. 

The Fig.~\ref{fig:spider-runtimes} is organized similarly to the one for Faida. The Fig.~\ref{fig:spider-memory} is the same, except it denotes the peak memory consumption.

The second experiment demonstrates run times and peak memory consumption depending on the maximum available memory. This experiment was run using the TPC-H dataset with the scale factor of 1. Its results are presented in Fig.~\ref{fig:spider-experiment}. This experiment is necessary to demonstrate how run time is affected when there is not enough memory.

\textbf{Results \& Discussion.}
Experiments demonstrated that the SETSTR implementation offers improvements ranging from 1.26 to 3.87 times (2.44 on average). VECPAIR and VECSV have shown almost identical results due to the similarity of these methods. When applying either of them, we achieve 3.92 improvement compared to Metanome on average, and 1.7 improvement compared to SETSTR. It is worth mentioning that for TPCH-10 both methods split \textit{lineitem} table in two chunks of the same size: VECPAIR splits it due to implementation cap on chunk size of 4GB, while VECSV hits memory limit and writes intermediate results to the disk. Therefore, \textit{lineitem} parts have to be merged on the later stages of both methods.

For some datasets, VECSTR shows the same run times compared to other techniques based on \verb+std::vector+, but it is usually outperformed by them due to different allocation mechanism used for \verb+std::string+ type. On the other hand, sorting routine is faster in VECSTR case.

Switching to memory consumption results depicted in Fig.~\ref{fig:spider-memory}, we can conclude that: (1) among proposed optimizations, SETSTR is the most efficient for datasets that contain large number of duplicates; (2) for other types of datasets applying VECPAIR would be a better choice; (3) VECSTR almost always leads to the largest peak memory consumption, as was presumed in Sec.~\ref{sec:implementation-faida}.

Note that analysis of memory consumption by Metanome and our versions of Spider can lead to some ambiguous conclusions. There are two reasons for that: (1) specifics of JVM and Garbage Collector (2) the way Metanome preprocesses data. For some datasets, Java implementations show significantly larger memory consumption, while for some Metanome yields the best results. E.g, HAINDGEN fully allocates into memory by Metanome and is processed in a column-wise manner, while our preprocessor splits dataset into ``wide'' chunks that contain all the table's attributes.

Our Spider implementations perform table processing in a one-pass manner, which explains larger memory consumption and leads to writing intermediate results to the disk. In Fig.~\ref{fig:spider-experiment}, we show run time as a function of the maximum available memory. The experiment was conducted on TPC-H dataset. We can draw three conclusions: (1) varying the parameter highly affects VECTORSTR: when setting 256 MB cap, the run time increases by a factor of 1.7; (2) when there is no need for VECPAIR to write to the disk (e.g. 2GB cap), it shows better results compared to VECSV and VECSTR; (3) starting with 512 MB memory limit, VECSV and VECSTR improve run time by a factor of 1.55 compared to SETSTR.

VECSV and VECPAIR achieve better performance on a multi-core setting compared to SETSTR. In regards to the run time, both methods are equally effective, though memory consumption rates differ. VECSV and VECPAIR perform well on datasets with size less or equal to 4GB, since there is no need for splitting data into chunks. However, one should prefer VECPAIR over VECSV as the former implementation requires much less memory. When dealing with datasets with size greater than 4GB VECSV should be preferred over VECPAIR, as the latter has to swap data chunks to the disk.

\subsection{Wrap-up and Takeouts}
Experiments demonstrated that it is possible to increase the performance of Faida and Spider by applying the proposed optimization techniques.

For Faida, we managed to obtain up to 8x improvement in terms of run times. But whats more important, is the fact that we proved that each of our techniques are useful since they all allow to yield positive results. The only questionable one is the SIMD-enabled execution, but we believe that switching to AVX-512 will unlock its full potential (recall the end of Sec.~\ref{sec:implementation-faida}). 

We were able to improve Spider's performance up to 5 times while reducing memory consumption. The obtained speed up of Spider comes from parallelization of the algorithm. Thus, increasing the number of threads (given the sufficient number of cores and table columns) may improve the results even further. Achieved improvements in Spider performance implies that it is also possible to speed up every spider-based algorithms, such as S-indd or S-indd++.

Having discussed the conducted experiments, we are now ready to summarize this section and give answers to the following research questions:

\begin{enumerate}
 \item \textit{RQ1: which families of optimization techniques can be considered as the more efficient for implementing automatic IND discovery algorithms?} For Faida-like algorithms, we conclude that all proposed optimization techniques can be considered useful, while the most efficient combination is \textit{parallelized version of techniques applied together} (number 6 on figures). For Spider-like algorithms we advise using vector-based implementations with varying memory limits: increase limits to reduce run time, or decrease it for smaller memory footprint.
 \item \textit{RQ2: how significantly can an engineering approach to the problem of automatic IND discovery improve state-of-the-art solutions?} Compared to the Metanome implementation, we were able to improve Faida run time by a factor ranging from 1.6 to 8.2 depending on the dataset. Improvement for Spider ranges from 1.26 to 4.89.
\end{enumerate}

\section{Threats to Validity}\label{sec:threats}

\begin{itemize}
    \item In the preprocessing phase, both algorithms use disk extensively for both reading and writing data. Therefore, their overall run time is very dependent on disk speed. In our experiments, we used a fast SSD. Run times may increase considerably for slow devices, e.g. HDD.
    \item Algorithm run times are affected by dataset properties: the number of dependencies, the number of unique values in a column, the number of columns, etc. In our experiments, we tried to evaluate several datasets with different properties. However, it is possible that we missed some cases.
    \item Running Metanome, we have not varied JVM parameters, except heap size in one of our experiments with Spider. It is possible that there is a combination of parameters that may increase the overall performance of an algorithm. However, our previous study~\cite{9435469} demonstrated that changing default parameters do not provide a substantial boost to performance.
    \item Finally, our experiments were conducted without varying Faida parameters. We performed our evaluation using values that were recommended by the authors of the original algorithm. Changing those values may affect the accuracy of the algorithm as well as its performance.
\end{itemize}

\section{Conclusion}\label{sec:conclusion}

In this paper, we have proposed a number of optimization techniques for two IND discovery algorithms: Faida and Spider. Experimental evaluation has demonstrated that we achieved improvement in both run time and memory consumption rate. Suggested optimizations can be applied to components of any other IND discovery solution with similar architecture.

\section*{Acknowledgment}

We would like to thank Vladislav Makeev for his help with the preparation of the paper. 

\bibliographystyle{IEEEtran}

\bibliography{bibliography-short-etal}

\end{document}